\documentclass[aps,prl,twocolumn,citeautoscript,showpacs,longbibliography]{revtex4}
\usepackage{graphicx}
\usepackage{caption}
\usepackage{subcaption}
\usepackage{amsmath}
\usepackage{amssymb}
\usepackage{dcolumn}
\usepackage{bm}

\begin{document}

\title{Fixed-node and fixed-phase approximations and their relationship to variable spins in quantum Monte Carlo}

\author{Cody A. Melton$^{1}$ and Lubos Mitas$^{1}$}
\affiliation{
1) Department of Physics, North Carolina State University, Raleigh, North Carolina 27695-8202, USA\\
}

\date{\today}

\begin{abstract}
 We compare the fixed-phase approximation with the
better known, but closely related fixed-node approximation on several testing examples. 
We found that both approximations behave very similarly with the fixed-phase results being very close to the fixed-node method whenever nodes/phase were of high and comparable accuracy. The fixed-phase exhibited larger biases when the trial wave functions errors in the nodes/phase were intentionally driven to unrealistically large values. We also present 
a formalism that enables to describe wave functions with the full antisymmetry in spin-spatial degrees of freedom using our recently developed method for systems with spins as fully quantum variables. This opens new possibilities for 
simulations of fermionic systems in the fixed-phase 
approximation formalism.
\end{abstract}

\maketitle

\subsection{Introduction}
Quantum Monte Carlo methods have proved to be very successful in calculations of many-body quantum systems. The number of applications as well as variety of algorithms is growing despite the fact that the fermion sign problem imposes a significant and fundamental challenge on the efficiency of stochastic approaches in general \cite{qmcrev,KolorencMitasReview}. In order to overcome this obstacle some type of approximation is introduced that avoids the inefficiences caused by the fermion signs and/or complex amplitudes. One of the most common approximations is the fixed-node and its closely related fixed-phase methods
\cite{anderson75,anderson76,reynolds82}. The fixed-node method
has been used now over four decades and it is an established approach that has led to a number of important calculations  that serve as benchmarks in comparisons with other approaches.
Although optimization of nodes is notoriously difficult, partial successes have been achieved such as using of parametrized effective Hamiltonian methods for generation of orbitals for
Slater-Jastrow wave functions \cite{kolorenc10}. On the other hand,
the fixed-phase approximation \cite{ortiz} is less familiar and much less established. Although it is known that fixed-node is a special case of the fixed-phase as was emphasized 
already in the original paper \cite{ortiz} and on occasions stated in other papers, it is fair to say that accurate data that would illustrate the behavior of these approximations side-by-side is scarce. 
The key point of this paper is to shed some new light
exactly on this aspect and to illustrate behavior of these approximations on some simple testing examples. 

\subsection{Fixed-phase approximation.} Let us consider the many-electron Hamiltonian $H=T+V$, where $V$ denotes electronic, ionic (local) and possibly other interactions and $T$ is the kinetic energy. When the desired eigenstate is real the stochastic methods of solitions are well-known and are mostly based on the fixed-node approximation as have been described in several reviews \cite{qmcrev,KolorencMitasReview}. Our focus this time is different and we assume that the state of a given symmetry that we are interested in is
- inherently or by construction - 
complex,  so that we can write 
$\Psi=\rho\exp(i\Phi)$ where $\rho({\bf R})\geq 0$ is a positive 
amplitude and $\Phi({\bf R})$ is a phase. We denote ${\bf R}=({\bf r}_1, ..., {\bf r}_N)$ for a set of coordinates of $N$ fermionic particles. If we substitute $\Psi$ into the imaginary-time 
 Schr\"odinger equation we get the following real and imaginary components
\begin{eqnarray}
-\partial_{\tau} \rho &=& [T+V+(\nabla\Phi)^2/2]\rho \\
-\partial_{\tau} \Phi &=&  [T\Phi - \rho^{-1}\nabla\rho.\nabla\Phi]
\end{eqnarray}
The imaginary part describes a conservation of 
the phase flow. The real part is actually the relation that 
provides the eigenvalue, ie, its solution  
 converges to the desired eigenstate in the limit $\lim_{t\to\infty} \rho(t)$.
We employ the 
machinery of projector quantum Monte Carlo methods that formally write the solution as a projection 
\begin{equation}
\rho_{ground}=\lim_{\tau\to\infty} \exp\{-\tau[T+V+(\nabla\Phi)^2/2]\}\rho_T
\end{equation}
where $\rho_T$
is an arbitrary positive amplitude and $\rho_{ground}$ is the ground state of the symmetry that is determined solely by the phase. Note that seemingly we have avoided the fermion sign problem since the amplitude that is to be sampled 
is non-negative everywhere. However, the source of the bias related to the avoidance of the fermion sign problem now becomes the potential term generated by the phase. Obviously, for a general eigenstate the phase is typically unknown and has to be approximated as outlined later. 
As it is also well-known, the fixed-phase is a special case of the fixed-node method, a simple demonstration can be found, for example, 
in \cite{melton2016b} and it is also demonstrated on an example that follows.

Before we formulate the fixed-phase approximation 
it is instructive to sketch a
simple problem with the phase. 
We will be testing the fixed-phase bias on an atomic $p$-state 
and since it is easy to create a complex version of such state
we will
use it for an illustration as well. Consider the following complex wave function
that is the  one-particle
ground state of $p-$symmetry for atomic Coulomb potential $V(r)=-1/r$
\begin{equation}
\psi=(x+icy)e^{-r/2}=r_{xy}h(r)\exp(i\Phi)
=\rho(r)\exp(i\Phi)
\end{equation}
where $c$ is a real constant. We have denoted 
$r_{xy} =\sqrt{x^2+c^2y^2}$, $h(r)=\exp(-r/2)$ and let us
remind that $r=\sqrt{x^2+y^2+z^2}$, in contrast with $r_{xy}$.
We find 
\begin{equation}
\Phi= \cot^{-1} (cy/x)
\end{equation}
and then we easily derive
\begin{equation}
\nabla\Phi= (cy/r_{xy}^2){\bf x}_0 -(cx/r_{xy}^2){\bf y}_0
\end{equation}
where unit vectors ${\bf x}_0, {\bf y}_0$ correspond to $x,y$ directions. The potential generated by the phase is given by
\begin{equation}
V_{ph}(c)=(1/2)(\nabla \Phi)^2 = { c^2 \over 2} {x^2+y^2 \over  r_{xy}^4}
\end{equation}
where $c$ plays a role of  a parameter.
It is straightforward to verify that the amplitude$\rho(r)=r_{xy}h(r)$ fulfills the Schrodinger equation  
\begin{equation}
\left[T-1/r +{c^2\over 2}{x^2+y^2)\over (x^2+c^2y^2)^2}\right]\rho(r)=E_{0p}\rho(r)
\end{equation}
where $E_{0p}=-1/8$ is the well-known hydrogenic eigenvalue for the $p-$state. 
It is clear that applying QMC methods to solve this equation would be straightforward. At the same time, it is equally straightforward to apply the fixed-node method, say, by using the trial function 
$\Psi_T=x\exp(-r/2)$ that has a node at the plane $x=0$ and to obtain an equivalent solution with the same eigenvalue, as had been done 
in the early days of QMC by Anderson \cite{anderson76}.
The example although seemingly trivial enables to illustrate the following two points. First, the fixed-node solution can be obtained 
by taking the limit $c\to\infty$ of the fixed-phase potential 
\begin{equation}
\lim_{c\to\infty} V_{ph}(c) = V_{\infty}\delta({\bf R}-{\bf R}_{\Gamma})
\end{equation}
where $V_{\infty}$ diverges as $const/c^2$ away from the origin and 
\begin{equation}
{\bf R}_{\Gamma}=\{{\bf R}; x=0\}
\end{equation}
so that the fixed-phase potential becomes a fixed-node potential that is more naturally understood as a zero value boundary condition applied on the wave function. The limit fixed-phase $\to$ fixed-node can be constructed completely generally as we have shown previously 
\cite{melton2016b}.
Second point is that we constructed a whole 
{\em manifold} of phase potentials  $V_{ph}(c)$ parametrized by $c$ for which the solution of the corresponding Schrodinger equation is 
{\em exact}. This is quite remarkable and shows that complex wave functions enable us to formulate the Schrodinger eigenvalue problem
in a somewhat new setting that might open new possibilities
for constructing approximations that avoid the fermion sign problem.

\begin{figure*}[!t]
    \centering
    \begin{subfigure}{0.3\textwidth}
        \centering
        \includegraphics[width=0.9\textwidth]{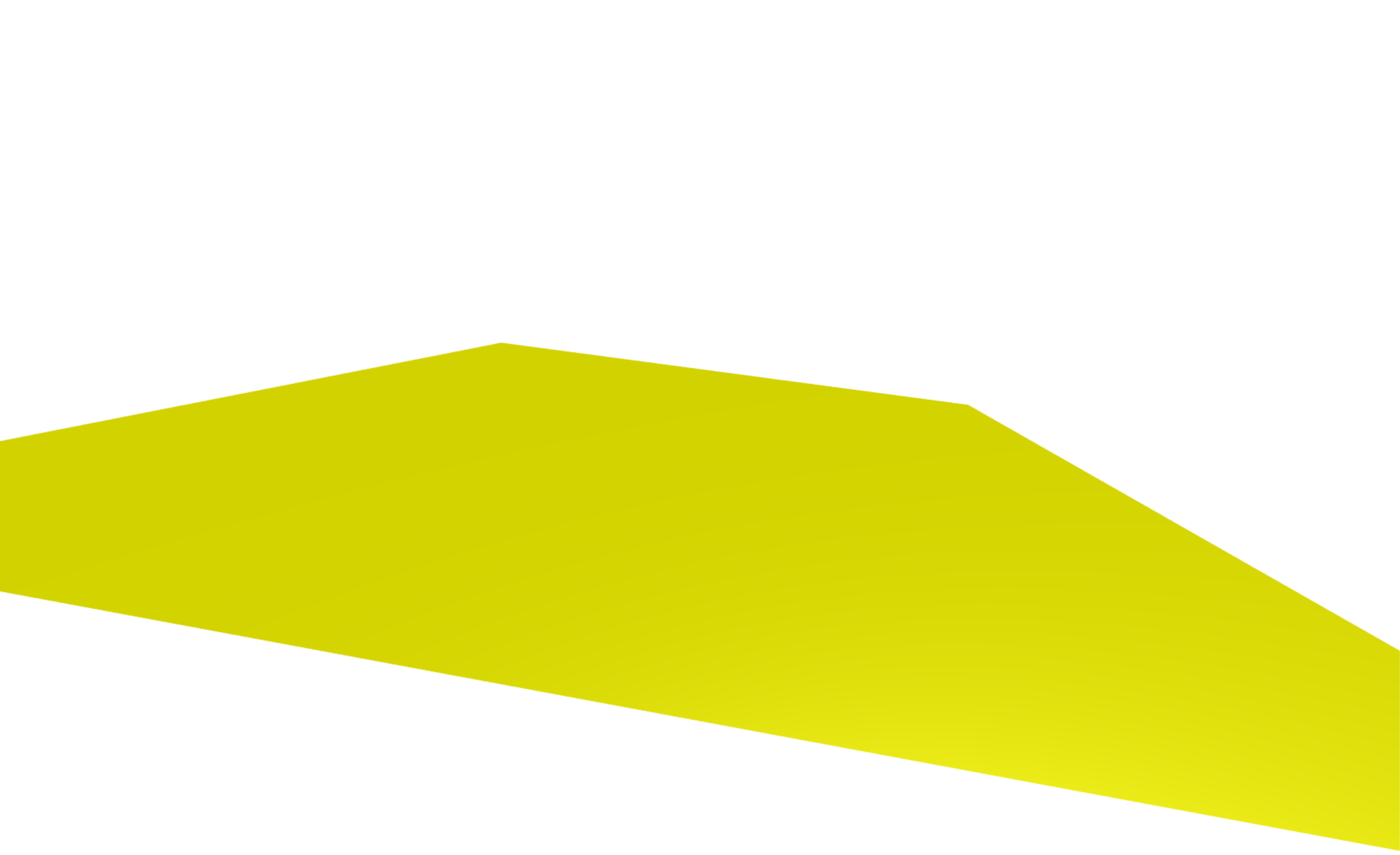}
        \caption{$\alpha = 0$, $\beta = 0$}
    \end{subfigure}
    \begin{subfigure}{0.3\textwidth}
        \centering
        \includegraphics[width=0.9\textwidth]{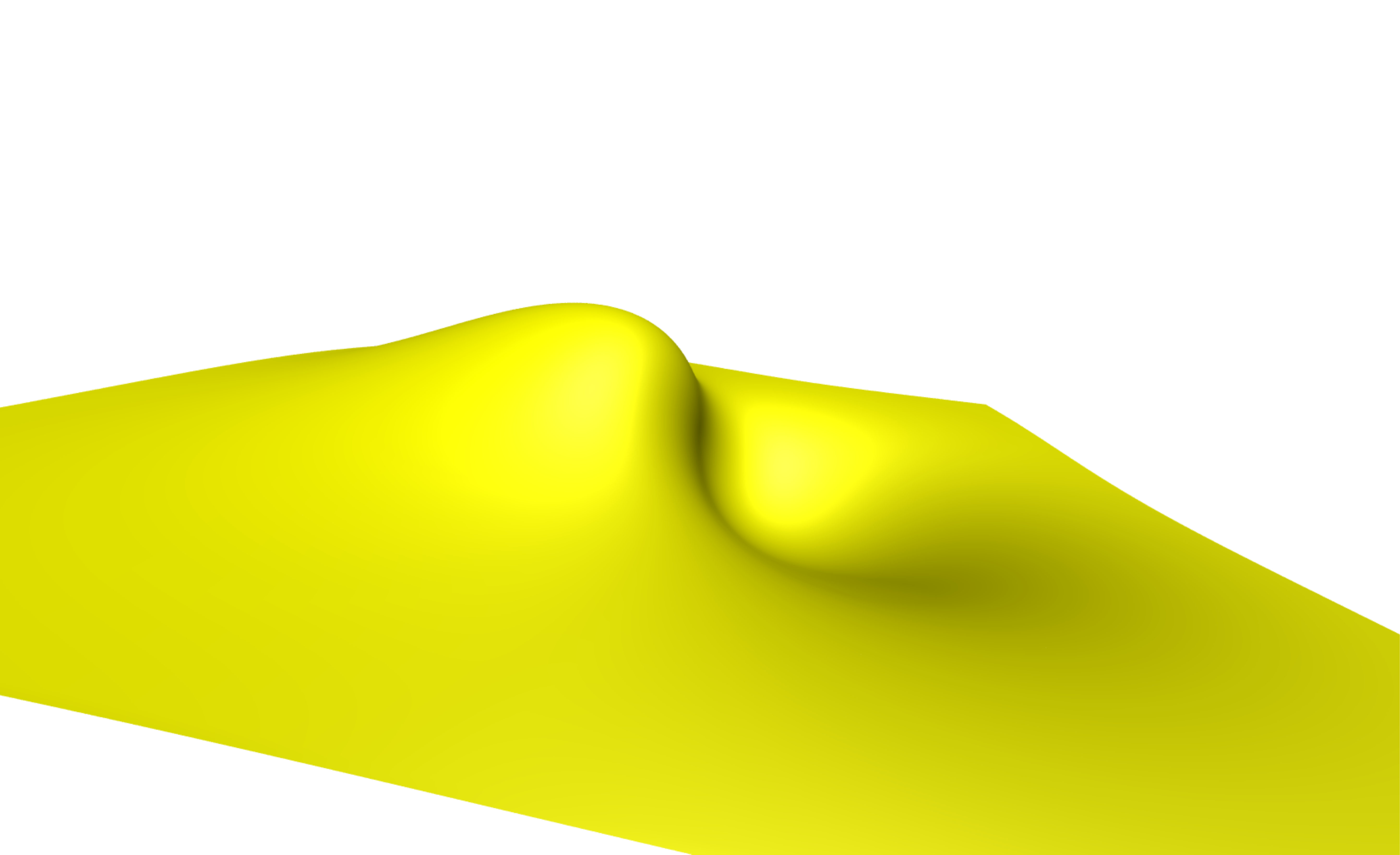}
        \caption{$\alpha = 1$, $\beta = 2$}
    \end{subfigure}
    \begin{subfigure}{0.3\textwidth}
        \centering
        \includegraphics[width=0.9\textwidth]{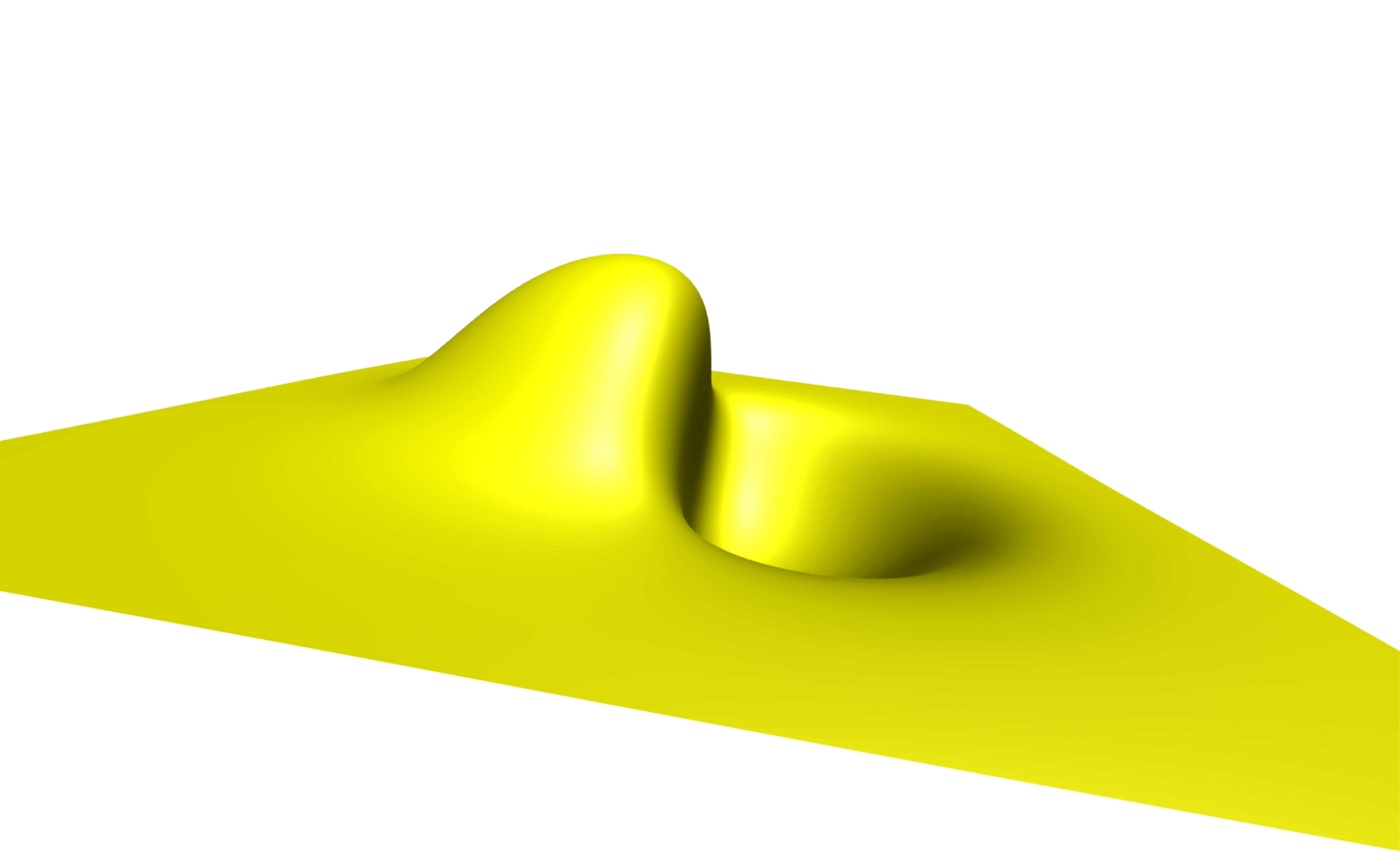}
        \caption{$\alpha = 1$, $\beta = 10$}
    \end{subfigure}
    \caption{Nodal surfaces for $\psi^T_C$ for various distortion parameters. Note that (b) is plotted for $x,y,z \in [-10,10]$~a.u. whereas (c) is plotted from $x,y,z \in [-2.5,2.5]$~a.u. The distortion is more localized at the origin for larger $\beta$.}
    \label{fig:nodes}
\end{figure*}

In fact, this fits our current purposes and we admit 
that we do not know the exact phase  
and we have to introduce some compromise. 
In the fixed-phase approximation
the exact phase is replaced by an appropriate
trial wave function phase \cite{ortiz}
so that the corresponding potential is given as 
\begin{equation}
V_{ph}= (\nabla\Phi)^2/2 \approx V_{ph,T}= (\nabla\Phi_T)^2/2
\end{equation}
where the trial function is $\Psi_T=\rho_T\exp(i\Phi_T)$.
The properties of the QMC method with approximate $V_{ph,T}$ are easy to understand. 
In particular, irregardless of  admissible $V_{ph,T}$, the method is variational as was stated very early on \cite{ortiz}.
This is easy to see since not only for $\rho_T$ but for {\em arbitrary} $\rho$ the variational wave function
$\Psi=\rho\exp(i\Phi_T)$ will lead to an upper bound to the exact energy that is given by 
\begin{equation}
E_{var}=\langle \Psi |(T+V)|\Psi \rangle=
\langle \rho |(T+V+V_{ph,T})|\rho\rangle.
\end{equation}
Clearly, the repulsive potential generated by an approximate
phase can only raise the total energy. The variational theorem also implies that the bias will be proportional to the square of the trial wave function error, therefore one expects that the fixed-phase method will have similar behavior as the more familiar fixed-node approach.
The question is whether the fixed-phase bias will not only have similar behavior but, more importantly, how large  the corresponding errors will be. In order to provide some insight into this question we will first study the following simple problem based on distortion of the exact node for the $p$-state as presented below.
Before we analyze this model we want to introduce the importance sampling for the fixed-phase method that is achieved by multiplying
the equation 
\begin{equation}
-\partial_{\tau} \rho = [T+V+(\nabla\Phi_T)^2/2]\rho
\end{equation}
with the trial amplitude $\rho_T$ and after arrangements we get
the following equation for the product $g=\rho\rho_T$ 
\begin{multline}
	\label{importance-sampling} -\frac{\partial g(\mathbf{R},\tau)}{\partial \tau} = -\frac{1}{2}\nabla^2 g(\mathbf{R},\tau) + \nabla \cdot \left[ \mathbf{v}_D(\mathbf{R}) g(\mathbf{R},\tau)\right] \\
    + \left[ E_L(\mathbf{R}) - E_T \right] g(\mathbf{R},\tau)
\end{multline}
where we have included an energy offset $E_T$. 
The importance sampling introduces the drift velocity 
\begin{equation}
\mathbf{v}_D(\mathbf{R}) = \nabla \ln  \rho_T(\mathbf{R}) = \rho_T^{-1}(\mathbf{R}) \nabla \rho_T(\mathbf{R})
\end{equation}
and the local energy 
\begin{equation}
  \label{eqn:local_energy}  E_L(\mathbf{R})  = \rho_T^{-1}(\mathbf{R}) \left[ -\frac{1}{2} \nabla^2 + V + \frac{1}{2}\left| \nabla \Phi_T(\mathbf{R})\right|^2 \right] \rho_T(R)
\end{equation}
These expressions are similar to the ones that appear in the fixed-node approach \cite{qmcrev} and can be solved by the same algorithm. 

\subsection{Simple model for comparing fixed-node vs. fixed-phase approximations}\label{section:FNvsFP}
Perhaps the simplest way how to study the differences between the fixed-node and fixed-phase approximations is consider a toy model with one electron subject to a central potential, namely 3D harmonic oscillator (HO) and Coulomb (C) potentials, with Hamiltonians
\begin{eqnarray}
    H_{HO} &=& -\frac{1}{2}\nabla^2 + \frac{1}{2} r^2 \\
    H_{C} &=& -\frac{1}{2}\nabla^2 - \frac{1}{r}
\end{eqnarray}

The ground state of these Hamiltonians are familiar, and are entirely nodeless. However, if we consider the first excited state of $p$ symmetry wiht the states being real, there is exactly one nodal plane. The eigenstates and eigenvalues (a.u.) for HO and C are respectively given by
\begin{eqnarray}
\psi_{HO}(r) &=&  z e^{-r^2/2}, \quad E = 5/2 \\
\psi_C(r) &=& z e^{-r/2},\quad  E = -1/8 
\end{eqnarray}
Since the exact eigenstates and nodal surface are known, any distortion to the nodal surface will yield the corresponding fixed-node error using the fixed-node DMC method. 
We therefore construct trial wave functions with distorted nodal surfaces of the form 
\begin{eqnarray}
    \label{distorted_HO}
    \psi_{HO}^T &=& c_0 z e^{-r^2/2} + \alpha c_1(\beta)x e^{-\beta r^2/2} \\
    \label{distorted_C}
    \psi_{C}^T &=& d_0 z e^{-r/2} + \alpha d_1(\beta) x e^{-\beta r/2}
\end{eqnarray}
where $\alpha$ and $\beta$ are the distortion parameters
while the normalization constants are given as $c_0=\sqrt{2/\pi^{3/2}}$, 
$c_1(\beta)=\sqrt{2\beta^{5/2}/\pi^{3/2}}$,
$d_0=\sqrt{1/32\pi}$, 
$d_1(\beta)=\sqrt{\beta^5/32\pi}$.
To illustrate the effect of the distortion parameters, nodal surfaces for several distortion parameters are shown in Fig. \ref{fig:nodes}.

In order to compare the fixed-node and fixed-phase errors of {\em equivalently} distorted wave functions, we construct complex versions of the wave functions in equations (\ref{distorted_HO}) \& (\ref{distorted_C}), namely
\begin{eqnarray}
    \begin{split}
       \psi_{HO}^T = c_0 z e^{-r^2/2} + \alpha c_1(\beta) x e^{-\beta r^2/2} \\ + i \left( c_0 y e^{-r^2/2} + \alpha c_1(\beta) x e^{-\beta r^2/2}\right) 
    \end{split}	\\
    \begin{split}
    \psi_{C}^T = d_0 z e^{-r/2} +  \alpha d_1(\beta) x e^{-\beta r/2} \\ + i\left(d_0 y e^{-r/2} +  \alpha d_1(\beta) x e^{-\beta r/2} \right)
\end{split}
\end{eqnarray}
which were used to construct trial amplitudes and phases. 

Since these wave functions are equivalent and should yield the same variational energy, we performed the variational Monte Carlo (VMC) calculations for each value of $\alpha$ and $\beta$ to ensure the energy was the same for the fixed-node (using $\psi_T$) and the fixed-phase (using $\rho_T$ and $\phi_T$). We then performed diffusion Monte Carlo (DMC) calculations for each system \cite{qmcrev,ortiz}. We compare the percentage error for both approximations and 
the results for both Hamiltonians are shown in Table I. 
For both Hamiltonians, we see excellent agreement at the VMC level, indicating that the trial wave functions are indeed equivalent. 

\begin{table*}[!t]
    \label{table:FNvsFP}
    \caption{Total energies (a.u.) of harmonic oscillator (HO) and Coulomb (C) systems in VMC,  fixed-node (FN) DMC and fixed-phase (FP) DMC methods, for various wave function distortions.
}
    \centering
    \begin{tabular}{|ccc|ccc|ccc|}
	\hline\hline
    &Hamiltonian &&& HO &&& C &\\
    \hline\hline
	$\alpha$ & $\beta$ & Method & VMC & DMC & $\%$ error & VMC & DMC & $\%$ error\\
	\hline
	0.1 & 1.1 & FN & 2.50011(2) & 2.50010(3) & 0.004(1)  & -0.124988(2) & -0.124987(4) & 0.009(3) \\
	    &     & FP & 2.50011(1) & 2.50010(2) & 0.0042(8) & -0.124987(1) & -0.124987(3) & 0.010(2)\\
	\hline
	& 2.0 & FN & 2.5062(1)  & 2.5038(1) & 0.153(5) & -0.12376(2) & -0.12405(5) & 0.75(4)\\
	&     & FP & 2.50620(9) & 2.5051(1) & 0.205(5) & -0.12376(2) & -0.12393(3) & 0.84(3) \\
	\hline 
	& 5.0 & FN & 2.5395(4) & 2.5208(5) & 0.83(2) & -0.1051(3) & -0.1205(2) & 3.5(2)\\
	&     & FP & 2.5396(3) & 2.5271(5) & 1.08(2) & -0.1051(4) & -0.1197(3) & 4.1(2)\\
	\hline
	& 10.0 & FN & 2.6001(8) & 2.5365(8) & 1.46(3) &  -0.024(4) & -0.1210(2) & 3.1(2) \\
	&      & FP & 2.6002(8) & 2.5457(6) & 1.82(2) &  -0.025(4) & -0.1203(3) & 3.6(2) \\
	\hline
	1.0 & 1.1 & FN & 2.5056(1) & 2.5041(2) & 0.16(1) & -0.12437(1) & -0.12442(3) & 0.46(2)\\
	    &     & FP & 2.5056(1) & 2.5046(2) & 0.18(1) & -0.12437(1) & -0.12441(4) & 0.47(3)\\
	\hline
	    & 2.0 & FN & 2.812(1)  & 2.585(2) & 3.40(9) & -0.0624(3) & -0.1143(5) & 8.4(4)\\
	    &     & FP & 2.8127(9) & 2.623(1) & 4.93(6) & -0.0625(3) & -0.1114(5) & 10.8(4) \\
	\hline
	    & 5.0 & FN & 4.500(6) & 2.734(3) & 9.3(1)  & 0.874(5) & -0.1070(8) & 14.3(6)\\
	    &     & FP & 4.499(5) & 2.821(3) & 12.8(1) & 0.875(5) & -0.1044(8) & 16.4(6) \\
	\hline
	    & 10.0 & FN & 7.56(2) & 2.719(2) & 8.79(8)  & 4.93(4) & -0.113(1) & 9.2(8) \\
	    &      & FP & 7.56(2) & 2.773(3) & 10.94(8) & 4.90(5) & -0.1122(9) & 10.2(7)\\
	\hline
    \end{tabular}
\end{table*}

Note that for larger node/phase distortions the fixed-phase bias is larger than the fixed-node bias although even for the largest distortions they remain of the same order of magnitude. The fact that the fixed-phase error is larger is not too difficult to understand since the corresponding repulsive potential $V_{ph}$
acts in the full $3N$-dimensional space of $N$ particles in 3D while the fixed-node shrinks into a $(3N-1)$-dimensional hypersurface. Although the vanishing of the wave function on this hypersurface distorts it in the whole space, 
the resulting bias appears to be smaller.

Note that although our model appears to be trivially of a one-particle type it is actually more general and reaches beyond the one-particle picture. In particular, for the harmonic oscillator it applies to a two-particle case with arbitrary interaction.  The reason is that in this case the exact nodal surface for the  
$p$-state is exactly known since the symmetry of HO enables to reformulate the Schrodinger equation in the center of mass and relative coordinates. In turn, this shows that the exact eigenstate for the two-particle $P$-symmetry triplet is
given analytically
\begin{equation}
\Psi_{exact} = (z_1-z_2) f(r_1,r_2,r_{12})
\end{equation}
where $f$ is a non-negative function
so that the exact node is given by $z_1=z_2$ \cite{taut}. Therefore, the study serves also as the simplest model for a two-particle interacting case.

\subsection{Real wave function recast into a complex form.}

After analyzing simple model it is interesting to ponder how the fixed-phase method would behave for a non-trivial interacting system with more than two particles so that electronic spin and corresponding symmetries enter the picture.
Perhaps the simplest system in this respect is the Li atom. This system has another advantage that its nodal surface is very well approximated by the single-reference Slater-Jastrow wave function
\begin{equation}
\Psi = {\rm det}^{\uparrow}[\phi_{1s},\phi_{2s}] \phi^{\downarrow}_{2s}\exp(U)
\end{equation}
 that provides accuracy better than 1 mHa for the total energy. Here we assume that the orbitals $\{\phi_i\},\{\phi_j\}$
are calculated in orbital theories such as Hartree-Fock or similar
and $U$ is an appropriate Jastrow factor.
In the extension of our arguments from the previous part we will use results from the very recent progress in treatment of the spin degrees of freedom in QMC \cite{melton16a, melton2016b}. In particular, for systems with spin-orbit interactions the wave function cannot be written in the above Slater-Jastrow form since the value (ie, orientation) of the spin varies; one has to write the wave function as an antisymmetric product of one-particle spinors. In general, the one-particle spinor is given as  
\begin{equation}
\chi({\bf r},s) =\phi^{\uparrow}({\bf r})\chi^{\uparrow}(s)+\phi^{\downarrow}({\bf r})\chi^{\downarrow}(s)
\end{equation}
where $\chi^{\uparrow,\downarrow}(s)$ are corresponding spin functions. 
The spin $s$ is treated as continuous (periodic) variable in the interval 
$(0,2\pi)$ and the spin functions are chosen as $\chi^{\uparrow}(s)=\exp(+ is)$, $\chi^{\downarrow}(s)=\exp(-is)$. The reasoning is further elaborated
in the mentioned papers. Using this representation enables one to exploit 
continuous sampling similarly to the usual spatial variables
and also much of the existing formalism for such calculations.
In this respect we can write the antisymmetric part of the trial wave function for the Li atom ground state $^2S(1s^22s)$ as follows 
\begin{equation}
\Psi_{T, anti}({\bf R, S}) = {\rm det}[\phi_{1s}\chi^{\uparrow}, \phi_{1s}\chi^{\downarrow},\phi_{2s}\chi^{\uparrow}]
\end{equation}
where ${\bf S}=(s_1,s_2,s_3)$ denotes the spin coordinates.
Note that unlike the fixed-node trial function, that is essentially a product of spin-up and -down terms, there is only
{\em one} determinant present here and it includes {\em both} spin channels. The full trial wave function $\Psi_T=\Psi_{T,anti} \exp(U)$ includes the Jastrow factor with electron-electron and
electron-ion correlation terms as further elaborated elsewhere
\cite{melton2016b}.

\begin{figure}
    \centering
    \caption{Total energy of the Li atom using the fixed-node medthod (FNDMC, constant full line), with error bar interval (dashed lines), compared with the fixed-phase, complex wave function (FPSODMC) formulation a function of the time step for spin degrees of freedom. The spatial coordinates were evolved with the time step $\tau_{spatial}=0.001$ a.u. Further details are explained in the text.}
    \label{fig:Li_atom}
    \includegraphics[width=0.49\textwidth]{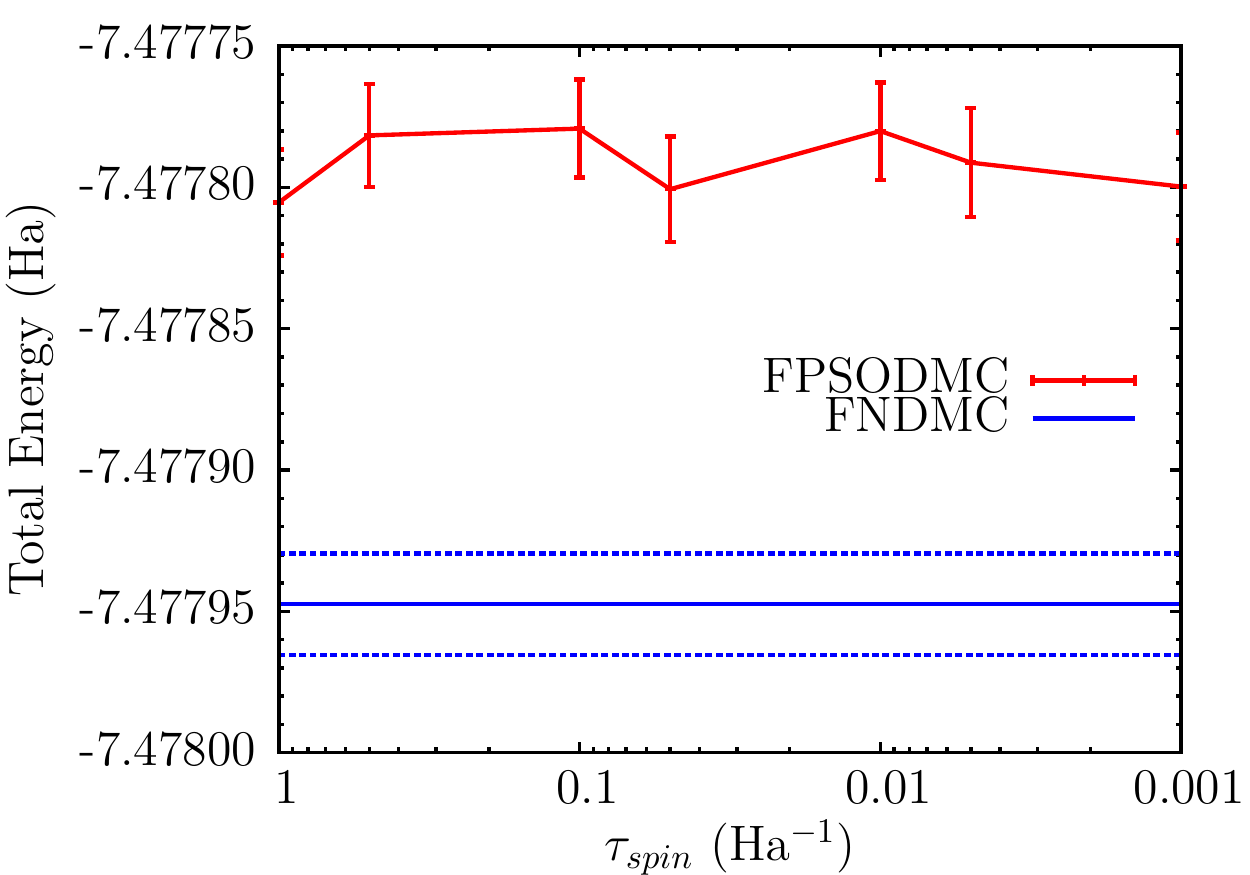}
\end{figure}

In effect, the wave function is a linear combination 
of three possibilities of how the spins up and down are assigned to the three particles. Let us define the following functions
\begin{equation}
\eta(s_1,s_2,s_3)=
e^{i(s_1+s_2-s_3)}
\end{equation}
and
\begin{equation}
D({\bf r}_1,{\bf r}_2, {\bf r}_3) ={\rm det}[\phi_{1s}({\bf r}_1),
\phi_{2s}({\bf r}_2)]\phi_{1s}({\bf r}_3)
\end{equation}
Then we can write the trial function as 
\begin{multline}
\Psi_T({\bf R,S})= \eta(s_1,s_2,s_3)D({\bf r}_1,{\bf r}_2, {\bf r}_3) \\ 
-\eta(s_2,s_3,s_1)D({\bf r}_2,{\bf r}_3, {\bf r}_1)
+\eta(s_3,s_2,s_1)D({\bf r}_3,{\bf r}_2, {\bf r}_1)
\end{multline}
so that it can be readily employed in the fixed-phase framework.
This is, in fact, the correct complete wave function for
the given state \cite{gill} as it takes into account the antisymmetry fully. Note that during the particle exchange, say, $1 \leftrightarrow 2$, {\em both} spatial {\em and} spin degrees of freedom are exchanged so that the symmetry of the wave function
should explicitly reflect that \cite{gill}.
Indeed, it is a combination of three possibilities how the spin projections can be distributed among the three electrons in this doublet state.
Clearly, more common is the fixed assignment of the spin (projection) to a given particle. That picks up one of the three spatial possibilities 
as it is routinely done in many approaches. Note that all three spatial determinants are equivalent, ie, related just by corresponding particle relabeling that does not affect the values of most of the expectations. Therefore both in quantum chemical calculations and also in QMC the focus is to calculate only the ``irreducible" spatial part, ie, one of the three possibilities that simplifies the problem so that it is enough to solve only for the spatial part of the wave function.
Note that it is indeed a combination of three possibilities how the spin for a doublet state can be distributed among three electrons. Consequently, it is also a linear combination of real wave functions that have different nodes corresponding to different particle spin assignments (``rotations") and
correspondingly different but equivalent spatial determinantal parts.  The linear prefactors are  complex and depend solely on the spins. In this manner we succeeded in complexifying the wave function so that fixed-phase method can be used to carry out the sampling and the whole calculation. It is obvious that now we explicitly consider both spatial and spin degrees of freedom
on the same footing. In order to evolve also the spin degrees of freedom we introduce a ``kinetic energy" operator that is given by
\begin{equation}
H_s({\bf S})=-(1/2\mu_s)\sum_i \left[ {\partial^2\over \partial s_i^2}+1\right]
\end{equation}
and it is added to the original Hamiltonian.
$H_s$ includes also energy offset so that this term does not 
contribute to the total energy. In actual calculations the effective spin mass $\mu_s$ plays a role of the spin time step that could be, in general, different from the spatial time step. Indeed this was how we carried out the calculations for the Li atom using the fixed-phase method with the trial function outlined above.
The one-particle orbitals were expanded in a gaussian basis with 
18 primitive functions. We checked that for our purposes the fixed-node calculation is very close to the fixed-node result with marginal difference of about 0.1 mHa due to the spatial time step bias. 
The Fig. \ref{fig:Li_atom} shows the results from the fixed-phase calculations with spin time step varying over three orders of magnitude compared with the usual fixed-node result. The important result is that there appears to be an increase in the bias from the fixed-phase formulation, however, 
it is very small of the order of 0.1 mHa, which is comparable to our overall time step bias that is already very marginal. Note that this fixed-phase bias seems to be uniform irregardless of the spin time step. Clearly, this shows that one can usefully reformulate the fixed-node setting into the fixed-phase framework without loosing any crucial accuracy. This opens interesting avenues to explore this method for other systems.

\subsection{Conclusions}
We have presented analysis and a few examples that compared the properties of the fixed-phase approximation with the
better known, but closely related, fixed-node approximation. 
For illustration we have chosen a few-particle systems that 
enabled analytic treatment and also allowed for using essentially exact fixed-node results as references. We found that both approximations behave similarly with the fixed-phase results being very close to the fixed-node method whenever nodes/phase were  of high and comparable
accuracy. The fixed-phase exhibited 
larger errors when distortions in the nodes/phase was intentionally driven to unrealistically large values. We have also presented 
a formalism that enables to describe wave functions with the full antisymmetry in spin-spatial degrees of freedom using our recently developed method for systems with significant 
spin-orbit interactions. This opens new possibilities for 
simulations of {\em any} fermionic system in the fixed-phase 
approximation formalism.

\subsection{Acknowledgments}

This research was supported by the U.S. Department of Energy (DOE), Office of Science, Basic Energy Sciences (BES) under Award DE-SC0012314. For calculations we used resources of the National Energy Research Scientific Computing Center, a DOE Office of Science User Facility supported by the Office of Science of the U.S. Department of Energy under Contract No. DE-AC02-05CH11231. We are very grateful for an
additional allocation at ANL Mira machine and for kind support from Dr. Anouar Benali. Part of the calculations have been carried out also at TACC.

\end{document}